\documentstyle[aps]{revtex}
\begin{document}

\title{Mechanism for long Dijkgraaf-Verlinde-Verlinde strings  }
\author{Vipul Periwal and \O yvind Tafjord}
\address{Department of Physics,
Princeton University,
Princeton, New Jersey 08544}

\def\dd{\hbox{d}}
\def\tr{\hbox{tr}}\def\Tr{\hbox{Tr}}
\def\ee#1{{\rm e}^{{#1}}}
%\baselineskip=14truept
\maketitle
\begin{abstract} 
The proposal of Dijkgraaf, Verlinde and Verlinde for the emergence of 
smooth strings from a supersymmetric $U(N)$ Yang-Mills theory  
assumes that conjugacy classes in the symmetric group 
with a few long cycles dominate the dynamics about the  infrared fixed 
point.  It is shown that the average number of cycles in a conjugacy
class of $S_{N}$ is bounded below
by  ${\rm const. }\sqrt 
N\ln\sqrt N,$ implying that some physical mechanism is needed to 
ensure the assumed dominance.  
It is shown that if individual cycles have positive energies that 
depend 
very weakly on their lengths,  then long cycles dominate the dynamics at
low temperatures.
 
\end{abstract}
Dijkgraaf, Verlinde and Verlinde\cite{dvv} have proposed a 
concrete model in which strings might appear from a two-dimensional
supersymmetric SU($N$) Yang-Mills theory, at an infrared fixed point in the
large $N$ limit, following the conjecture of 
Banks, Fischler, Shenker and Susskind\cite{bfss}, and using the 
result of Taylor\cite{wati}.  Their construction involves
identification of cycles in conjugacy classes 
 in the Weyl group of SU($N$), the permutation group on $N$ letters, 
 $S_{N},$ with string configurations.
 An important assumption in their
construction is that the dominant  physical configurations are those
corresponding to conjugacy classes with a few 
long cycles.  We wish in this paper to analyze this
assumption, and to suggest a physical analogy for  their infrared fixed 
point, which may help explain the assumed dominance of long cycles.

We start by considering the average number of cycles in a given 
permutation of the symmetry group on $N$ letters, $S_{N}$.  The total 
number of conjugacy classes in $S_{N}$ is, of course, $p(N),$ the
partition function that counts the number of ways that a positive integer 
$N$ may be written as a sum  of positive integers.  As is well known,
\begin{equation}
\sum p(n) t^{n} \equiv \prod (1-t^{i})^{-1}. 
\end{equation}
To calculate the average number of cycles in a conjugacy class,
we consider the modified generating function 
\begin{equation}
F(t,x)\equiv \sum p(n,m) t^{n} x^{m} \equiv \prod (1-xt^{i})^{-1}. 
\end{equation}
Clearly $p(n,m)$ is the number of partitions of $n$ into
$m$ cycles.  We want to compute the average 
$\sum_{m} mp(n,m)/p(n).$

\def\ggt{\gg}
\def\part{\partial}
To this end, consider 
\begin{equation}
x\part_{x}F = \sum mp(n,m) t^{n} x^{m} = F \sum {xt^{n}\over1-xt^{n}} ,
\end{equation}
so setting $x=1,$ and using
\begin{equation}
{1\over n} {t^{n}\over1-t} <{t^{n}\over1- t^{n}}<  {1\over n} {t 
\over1-t},
\end{equation}
we derive 
\begin{equation}
\part_{x}F(t,x=1) > -F(t,x=1) {\ln(1-t)\over1-t}.
\label{collect}
\end{equation}
Now, according to Rademacher\cite{rad},
\begin{equation} 
p(n) \asymp {{\exp(K\sqrt n)}\over {4n\sqrt 3}},
\end{equation}
with $K = \pi \sqrt (2/3).$
We need to compare the coefficients of $t^{n}$ on the two sides of
eq.~\ref{collect}.  On the right hand side, we find that the 
coefficient of $t^{n}$ is (asymptotically) greater than
\begin{equation}
 p(n) \int^{\sqrt n}_{0} dx \sqrt n\ln\sqrt n \exp\left(-{Kx\over 2}\right)
\approx \ { 2\over K} p(n) \sqrt n\ln\sqrt n.
\end{equation} 
Thus  
\begin{equation}
\langle m\rangle_{n}\equiv {{\sum_{m} mp(n,m)}\over p(n)} > O(\sqrt{n} \ln\sqrt n).
\end{equation}
We see that the average number of cycles in a conjugacy class grows quite 
rapidly  with $n.$  For the physical problem at hand, there must
be some dynamical mechanism that results in the matrix model behaving 
as a model of smooth strings near the infrared fixed point.
This is the issue we now 
address.

Recall that in the proposal of Dijkgraaf, Verlinde and 
Verlinde\cite{dvv}, 
turning on the string coupling implies going away from the infrared 
fixed point.  Thus it is not enough for the long cycles to dominate 
only at the infrared fixed point, they must be present even slightly 
away from the critical point.  It is important therefore that the 
physics that leads to the suppression of states with many cycles 
should be independent of the string coupling.  Further, the 
interactions, as defined in \cite{dvv}, are independent of the
lengths of cycles, so the mechanism for suppressing configurations 
with short cycles cannot rely on the nature of the interactions.

There is a well-known statistical mechanics paradigm for this 
situation.  Consider a two state system, with states labelled $A$ and
$B,$ with entropies such that 
$S_{A}\ggt S_{B}$ and energies such that $E_{A}\ggt E_{B}.$  Then the
probability of being in state $B$ is given by
\begin{equation}
P_{B} \equiv {1\over {1+\exp\left(\beta (E_{B}-E_{A})  - 
(S_{B}-S_{A})\right)}}.
\label{prob}
\end{equation}
Computing $\beta_{{1\over 2}},$ the inverse temperature at which
$P_{B}= 1/2,$ we find
\begin{equation}
\beta_{{1\over 2}} \ = {{S_{B}-S_{A}}\over E_{B}-E_{A}},
\end{equation}
and 
\begin{equation}
\beta_{{1\over 2}} {{\partial P_{B}}\over \partial\beta}({\beta_{{1\over 
2}}}) =   {1\over  4} (S_{A} - S_{B}).
\end{equation}
Thus, there is a very sharp transition from state $A$ begin occupied to
state $B$ being occupied as the temperature is decreased.

In the case at hand, state $A$ corresponds to the conjugacy classes 
with about $\sqrt N \ln \sqrt N$ cycles, and state $B$ to conjugacy classes with
very few cycles, with $S_{A} \approx  \ln p(N),$ and $S_{B}\approx O(1).$
From the heuristic description of the states given by Dijkgraaf,
Verlinde and Verlinde\cite{dvv}, it seems reasonable to suppose that the 
energy of a cycle should depend only on its length.  Further, we 
assume that $\beta_{1\over 2}$ should be finite, since if this 
inverse transition temperature were to go to zero, long cycles would
dominate the dynamics at all values of the string coupling, 
including where the underlying Yang-Mills 
theory would be weakly coupled\cite{dvv}. 
For a finite value of $\beta_{1\over 2},$ we must then have 
\begin{equation}
\sqrt N \ln \sqrt N E(\sqrt N/\ln\sqrt N) \ggt E(N),
\end{equation}
where $E(i)$ is the energy of a cycle of length $i.$  This implies 
  that $E(i) = 
{\rm const.}\ i +\epsilon(i),$ and 
$\epsilon(N) \approx \ln \sqrt N \epsilon(\sqrt N/\ln\sqrt N)$
for a finite transition temperature.  (The term proportional to
the length of the cycle is a measure of the number of consituents
in each cycle, so it is not relevant for comparing
configurations for a fixed value of $N.$) 
 We can conclude that   $\epsilon(i)$
should be approximately  constant.    
(A very weak  dependence on $i,$ {\it e.g.} about $\ln i,$ cannot be
ruled out by our simple estimate.)  
Finally, 
consider the sign of $\epsilon(i):$ For this mechanism to work, 
$\epsilon$ must  be {\it positive}.  A speculation is that $\epsilon$
is   a sort of Casimir energy for each cycle, sensitive only to the
periodicity that defines each cycle.  Given its weak dependence on the
number of constituents, it seems difficult to interpret $\epsilon$ as
an interaction energy.  

In conclusion, we have calculated a lower bound for
the average number of cycles in a
conjugacy class of the Weyl group of $U(N),$ $S_{N}.$  This number 
grows quite rapidly with $N,$ so we suggested a simple physical 
mechanism   which would imply the assumed dominance\cite{dvv} by conjugacy 
classes with a few long cycles.

This work was supported in part by NSF grant PHY96-00258.

\end{document}